\documentclass[apj]{emulateapj}
\usepackage{times}
\usepackage{bm}
\usepackage{ifthen}

\usepackage{threeparttable}

\newcommand{\sk}{Super-K~}

\newcommand{\degree}{$^\circ$}

\newboolean{ApJnames}
\setboolean{ApJnames}{false}

\ifthenelse{\boolean{ApJnames}}{
}
{
\graphicspath{{./figs/}}
}

\begin{document}
\newcommand{\rmk}[1]{\emph{#1}}

 \shorttitle{Neutrino Astronomy with Super-Kamiokande-I}

 \shortauthors{K. Abe et al.}

\title{High energy neutrino astronomy using upward-going muons in
  Super-Kamiokande-I }

\newcounter{foots}
\newcounter{notes}
\newcommand{\authoraticrr}{$^{1}$}
\newcommand{\authoratncen}{$^{2}$}
\newcommand{\authoratbu}{$^{3}$}
\newcommand{\authoratbupenn}{$^{3,\dagger}$}
\newcommand{\authoratbnl}{$^{4}$}
\newcommand{\authoratuci}{$^{5}$}
\newcommand{\authoratcsu}{$^{6}$}
\newcommand{\authoratcnu}{$^{7}$}
\newcommand{\authoratduke}{$^{8}$}
\newcommand{\authoratgmu}{$^{9}$}
\newcommand{\authoratgifu}{$^{10}$}
\newcommand{\authoratuh}{$^{11}$}
\newcommand{\authoratui}{$^{12}$}
\newcommand{\authoratkek}{$^{13}$}
\newcommand{\authoratkekicrr}{$^{13,1}$}
\newcommand{\authoratkekkashiwa}{$^{13,\ddagger}$}
\newcommand{\authoratkobe}{$^{14}$}
\newcommand{\authoratkyoto}{$^{15}$}
\newcommand{\authoratkyototriumf}{$^{15,\S}$}
\newcommand{\authoratlanluci}{$^{16,5}$}
\newcommand{\authoratlsu}{$^{17}$}
\newcommand{\authoratumd}{$^{18}$}
\newcommand{\authoratmit}{$^{19}$}
\newcommand{\authoratduluth}{$^{20}$}
\newcommand{\authoratmiyagi}{$^{21}$}
\newcommand{\authoratnagoya}{$^{22}$}
\newcommand{\authoratsuny}{$^{23}$}
\newcommand{\authoratniigata}{$^{24}$}
\newcommand{\authoratokayama}{$^{25}$}
\newcommand{\authoratosaka}{$^{26}$}
\newcommand{\authoratseoul}{$^{27}$}
\newcommand{\authoratshizuoka}{$^{28}$}
\newcommand{\authoratshizuokaseika}{$^{29}$}
\newcommand{\authoratskku}{$^{30}$}
\newcommand{\authorattohoku}{$^{31}$}
\newcommand{\authorattokai}{$^{32}$}
\newcommand{\authorattit}{$^{33}$}
\newcommand{\authorattokyo}{$^{34}$}
\newcommand{\authoratwarsaw}{$^{35}$}
\newcommand{\authoratwarsawuci}{$^{35,5}$}
\newcommand{\authoratuw}{$^{36}$}
\newcommand{\authoratuwduluth}{$^{36,20}$}
\newcommand{\addressoficrr}[1]{$^{1}$ #1 }
\newcommand{\addressofncen}[1]{$^{2}$ #1 }
\newcommand{\addressofbu}[1]{$^{3}$ #1 }
\newcommand{\addressofbnl}[1]{$^{4}$ #1 }
\newcommand{\addressofuci}[1]{$^{5}$ #1 }
\newcommand{\addressofcsu}[1]{$^{6}$ #1 }
\newcommand{\addressofcnu}[1]{$^{7}$ #1 }
\newcommand{\addressofduke}[1]{$^{8}$ #1 }
\newcommand{\addressofgmu}[1]{$^{9}$ #1 }
\newcommand{\addressofgifu}[1]{$^{10}$ #1 }
\newcommand{\addressofuh}[1]{$^{11}$ #1 }
\newcommand{\addressofui}[1]{$^{12}$ #1 }
\newcommand{\addressofkek}[1]{$^{13}$ #1 }
\newcommand{\addressofkobe}[1]{$^{14}$ #1 }
\newcommand{\addressofkyoto}[1]{$^{15}$ #1 }
\newcommand{\addressoflanl}[1]{$^{16}$ #1 }
\newcommand{\addressoflsu}[1]{$^{17}$ #1 }
\newcommand{\addressofumd}[1]{$^{18}$ #1 }
\newcommand{\addressofmit}[1]{$^{19}$ #1 }
\newcommand{\addressofduluth}[1]{$^{20}$ #1 }
\newcommand{\addressofmiyagi}[1]{$^{21}$ #1 }
\newcommand{\addressofnagoya}[1]{$^{22}$ #1 }
\newcommand{\addressofsuny}[1]{$^{23}$ #1 }
\newcommand{\addressofniigata}[1]{$^{24}$ #1 }
\newcommand{\addressofokayama}[1]{$^{25}$ #1 }
\newcommand{\addressofosaka}[1]{$^{26}$ #1 }
\newcommand{\addressofseoul}[1]{$^{27}$ #1 }
\newcommand{\addressofshizuoka}[1]{$^{28}$ #1 }
\newcommand{\addressofshizuokaseika}[1]{$^{29}$ #1 }
\newcommand{\addressofskku}[1]{$^{30}$ #1 }
\newcommand{\addressoftohoku}[1]{$^{31}$ #1 }
\newcommand{\addressoftokai}[1]{$^{32}$ #1 }
\newcommand{\addressoftit}[1]{$^{33}$ #1 }
\newcommand{\addressoftokyo}[1]{$^{34}$ #1 }
\newcommand{\addressofwarsaw}[1]{$^{35}$ #1 }
\newcommand{\addressofuw}[1]{$^{36}$ #1 }
\def\pennnow{$\dagger$}
\def\kashiwanow{$\ddagger$}
\def\triumfnow{\S}
\author{
{\bf The Super-Kamiokande Collaboration} \\
\vspace{0.2cm}
K.~Abe\authoraticrr,
J.~Hosaka\authoraticrr,
T.~Iida\authoraticrr,
K.~Ishihara\authoraticrr,
J.~Kameda\authoraticrr,
Y.~Koshio\authoraticrr,
A.~Minamino\authoraticrr,
C.~Mitsuda\authoraticrr,
M.~Miura\authoraticrr,
S.~Moriyama\authoraticrr
% ...
\\
\smallskip
\smallskip
\footnotesize
\it
\addressoficrr{Kamioka Observatory, Institute for Cosmic Ray Research, 
University of Tokyo, Kamioka, Gifu, 506-1205, Japan}\\
\addressofncen{Research Center for Cosmic Neutrinos, Institute for Cosmic 
Ray Research, University of Tokyo, Kashiwa, Chiba 277-8582, Japan}\\
\addressofbu{Department of Physics, Boston University, Boston, MA 02215, 
USA}\\
\addressofbnl{Physics Department, Brookhaven National Laboratory, Upton, 
NY 11973, USA}\\
\addressofuci{Department of Physics and Astronomy, University of 
California, Irvine, Irvine, CA 92697-4575, USA }\\
\addressofcsu{Department of Physics, California State University, 
Dominguez Hills, Carson, CA 90747, USA}\\
\addressofcnu{Department of Physics, Chonnam National University, Kwangju 
500-757, Korea}\\
\addressofduke{Department of Physics, Duke University, Durham, NC 27708, 
USA} \\
\addressofgmu{Department of Physics, George Mason University, Fairfax, VA 
22030, USA }\\
\addressofgifu{Department of Physics, Gifu University, Gifu, Gifu 
501-1193, Japan}\\
\addressofuh{Department of Physics and Astronomy, University of Hawaii, 
Honolulu, HI 96822, USA}\\
\addressofui{Department of Physics, Indiana University, Bloomington,
  IN 47405-7105, USA} \\
\addressofkek{High Energy Accelerator Research Organization (KEK), 
Tsukuba, Ibaraki 305-0801, Japan }\\
\addressofkobe{Department of Physics, Kobe University, Kobe, Hyogo 
657-8501, Japan}\\
\addressofkyoto{Department of Physics, Kyoto University, Kyoto 606-8502, 
Japan}\\
\addressoflanl{Physics Division, P-23, Los Alamos National Laboratory, Los 
Alamos, NM 87544, USA }\\
\addressoflsu{Department of Physics and Astronomy, Louisiana State 
University, Baton Rouge, LA 70803, USA }\\
\addressofumd{Department of Physics, University of Maryland, College Park, 
MD 20742, USA }\\
\addressofmit{Department of Physics, Massachusetts Institute of 
Technology, Cambridge, MA 02139, USA}\\
\addressofduluth{Department of Physics, University of Minnesota, Duluth, 
MN 55812-2496, USA}\\
\addressofmiyagi{Department of Physics, Miyagi University of Education, 
Sendai, Miyagi 980-0845, Japan}\\
\addressofnagoya{Solar Terrestrial Environment Laboratory, Nagoya University, Nagoya, Aichi 
464-8602, Japan}\\
\addressofsuny{Department of Physics and Astronomy, State University of 
New York, Stony Brook, NY 11794-3800, USA}\\
\addressofniigata{Department of Physics, Niigata University, Niigata, 
Niigata 950-2181, Japan }\\
\addressofokayama{Department of Physics, Okayama University, Okayama, 
Okayama 700-8530, Japan} \\
\addressofosaka{Department of Physics, Osaka University, Toyonaka, Osaka 
560-0043, Japan}\\
\addressofseoul{Department of Physics, Seoul National University, Seoul 
151-742, Korea}\\
\addressofshizuoka{Department of Systems Engineering, Shizuoka University, 
Hamamatsu, Shizuoka 432-8561, Japan}\\
\addressofshizuokaseika{Department of Informatics in Social Welfare, Shizuoka University 
of Welfare, Yaizu, Shizuoka, 425-8611, Japan}\\
\addressofskku{Department of Physics, Sungkyunkwan University, Suwon 
440-746, Korea}\\
\addressoftohoku{Research Center for Neutrino Science, Tohoku University, 
Sendai, Miyagi 980-8578, Japan}\\
\addressoftokai{Department of Physics, Tokai University, Hiratsuka, 
Kanagawa 259-1292, Japan}\\
\addressoftit{Department of Physics, Tokyo Institute for Technology, 
Meguro, Tokyo 152-8551, Japan }\\
\addressoftokyo{The University of Tokyo, Tokyo 113-0033, Japan }\\
\addressofwarsaw{Institute of Experimental Physics, Warsaw University, 
00-681 Warsaw, Poland }\\
\addressofuw{Department of Physics, University of Washington, Seattle, WA 
98195-1560, USA}
}

\altaffiltext{\pennnow}{Present address: Center for Gravitational Wave Physics, Pennsylvania State 
University, University Park, PA 16802, USA}
\altaffiltext{\kashiwanow}{Present address: Institute for Cosmic Ray Research, University of Tokyo, Kashiwa, Chiba, 
277-8582, Japan}
\altaffiltext{\triumfnow}{Present address: TRIUMF, Vancouver, British Columbia V6T 2A3, Canada}

\slugcomment{Submitted to ApJ on June 16, 2006}

\begin{abstract}

  We present the results from several studies used to search for
  astrophysical sources of high-energy neutrinos using the
  Super-Kamiokande-I (April~1996 to July~2001) neutrino-induced
  upward-going muon data.  The data set consists of 2359 events with
  minimum energy 1.6~GeV, of which 1892 are through-going and 467 stop
  within the detector.  The results of several independent analyses are
  presented, including searches for point sources using directional and
  temporal information and a search for signatures of cosmic-ray
  interactions with the interstellar medium in the upward-going muons.
  No statistically significant evidence for point sources or any diffuse
  flux from the plane of the galaxy was found, so specific limits on
  fluxes from likely point sources are calculated.  The 90\%~C.L. upper
  limits on upward-going muon flux from astronomical sources which are
  located in the southern hemisphere and always under the horizon for
  Super-Kamiokande are $(1 \sim 4)\times
  10^{-15}\rm{cm}^{-2}\rm{s}^{-1}$.

\end{abstract}

\maketitle


\section{Introduction} 
\label{sec:intro}

There have been attempts to find extraterrestrial neutrino sources since
the first underground experiments to observe cosmic-ray
neutrinos~\citep{Reines65,Achar65}.  Only two such sources have been
observed so far, MeV electron neutrinos from the
Sun~\citep{davis03,koshiba03} and from
SN1987A~\citep{Hirata87,Bionta87}.

At GeV and higher energies, the observational method involves analyzing
the arrival directions of neutrino-induced muons traversing underground
detectors, since muons over about 1~GeV travel in roughly the same
direction as their parent neutrinos.  It is desirable to observe as high
an energy sample of neutrinos as possible, as not only does this
pointing accuracy improve with energy, but astrophysical
signals are expected to have harder spectra than atmospheric neutrinos,
gaining in signal-to-noise ratio in direct proportion to the energy
threshold~\citep{Gaisser95}.  Muons above a few GeV cannot be contained in 
the detector,
so we restrict our search to only upward-going muons (UGMs) entering the
detector.  Such muons may be produced some distance away from the active
detector volume, by charged current interactions of muon neutrinos (and
anti-neutrinos) in the rock surrounding the detector.  The effective
volume for such instruments is thus roughly equal to the muon detection
area times the range of the muon in the surrounding rock.  The
combination of these factors means that any astrophysical point source
of neutrinos found by this analysis would in the TeV energy range.

Neutrinos may come from either astrophysical sources or cosmic ray
interactions with Earth's atmosphere or the interstellar medium.
Atmospheric neutrinos led to the observation of muon neutrino
oscillations~\citep{Fukuda98} but constitute an irreducible background
in the search for astrophysical neutrinos.  Furthermore, muons produced
in the atmosphere penetrate from the surface at rates greater than muons
from neutrino interactions in the surrounding rock, even in the deepest
of mines. However, this background can be largely removed by considering
only muons arriving from below the horizon, which can only be due to
neutrino interactions.

Given that cosmic rays are highly accelerated hadrons and neutrinos are
produced as decay products of energetic hadron interactions, it follows
that substantial neutrino fluxes must exist as well, although with
orders of magnitude uncertainty in fluence.  Many studies carried out
over the last 25 years agree that a detector on the order of
$10^6~\rm{m}^2$ muon collection area would be required to
see such sources~\citep{Learned00} if they are at the lower end of the
likely range of fluence.  Thus, at the sensitivity level of
Super-Kamiokande (``\sk'') (about 1200~$\rm{m}^2$ muon collecting area),
the chance to unambiguously identify a source may be relatively small
but certainly not negligible.  Additionally, bursting, highly beamed
like objects such as gamma-ray bursts and blazars or other transient
X-ray/$\gamma$-ray sources for which the main mechanism is unknown and
neutrino flux predictions are highly uncertain, may produce fluxes of
neutrinos detectable by \sk.


We report on analyses of the first five years of \sk UGM data, searching
for point sources and for diffuse flux from the galactic plane. No
significant evidence for high energy neutrino sources has been found. We
report upper limits on neutrino-induced UGM fluxes from the directions
of known high energy gamma ray sources, which in many cases are lower
than those previously set.


\section{The Super-Kamiokande Detector}
\label{sec:detector}

\sk is a 50~kton ring imaging water Cherenkov detector.  Results
presented here are derived from the ``SK-I'' running period, which began
when \sk started taking data in April~1996 and ended when it was shut 
down for maintenance in
July~2001.  

The detector is located in the Mozumi mine of the Kamioka Mining and
Smelting Company, in Gifu prefecture, Japan, directly underneath the
peak of Mt.~Ikenoyama with a minimum overburden of 2,700~m water
equivalent.  Its geodetic location is at 36.4\degree~N, 
137.3\degree~E, 
and altitude 370~m.  When looking at UGMs, \sk's northern location makes it sensitive to source
locations on the celestial sphere with declination ($\delta$) less than $54^\circ$.

The inner detector volume is a cylinder of radius 16.9~m and
height 36.2~m, viewed during SK-I by 11,146 50~cm diameter inward-facing
Hamamatsu photo-multiplier tubes (PMTs).  Approximately 40\% of the
inner detector surface area is photo-cathode, with the remainder covered
by light-absorbing black plastic. The outer detector volume, completely
surrounding and optically isolated from the inner detector, is a layer
of water 2.5~m thick viewed by 1,885 outward-facing 20~cm PMTs.  
A more detailed description of the detector can be found 
in~\citet{Fukuda03}.

By using amplitude and timing information from the PMT signals, muon
tracks in the inner detector can be reconstructed with tens of
centimeters accuracy for endpoints and a few degrees accuracy for
direction, depending upon particle momentum~\citep{Fukuda03,Desai04a,Ashie05}.
The outer detector is used to identify particles
entering and exiting the inner detector.



\section{Data and Background}
\label{sec:dataandbackground}

\subsection{Data}
\label{sec:data}

The dataset for these studies consists of all UGMs, both through-going
and those that stopped within the inner detector.  Contamination from cosmic ray
muons prohibits the use of downward-going muons.  Events were required
to have path length in the detector greater than 7~m (corresponding to a
muon energy greater than 1.6~GeV) to ensure good reconstruction
accuracy.  Parent neutrino energies, as determined by Monte Carlo
simulation, are broadly distributed, with peaks around 10~GeV for
stopping muons and 100~GeV for through-going muons. Further details on
the data filtering and reconstruction may be found
elsewhere~\citep{Ashie05}.

Events with deposited energy greater than 1,750,000 equivalent
photoelectrons (approximately 200~GeV deposited in the detector) were cut
due to the likelihood of phototube saturation. A separate study has been
done using those events~\citep{Swanson06}.  A $dE/dx$ based method of
tagging higher energy UGMs is discussed in~\citet{Desai04a}, but
is not used in this paper's analysis.  

\begin{figure}[htbp]
\begin{center}
\ifthenelse{\boolean{ApJnames}}
  {\plotone{f1.eps}}
  {\plotone{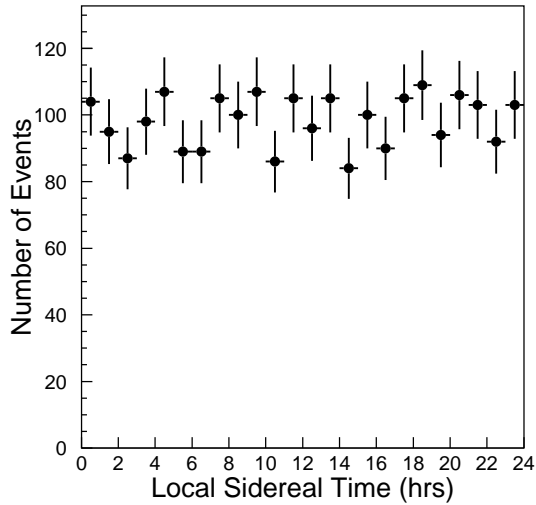}}
\caption[sidertime]{Sidereal time distribution of observed upward-going 
muons. There are no significant deviations from uniformity.}
\label{fig:sidereal}
\end{center}
\end{figure}

The total UGM livetime for the SK-I dataset was 1679.59 days, about 87\% of
calendar time over the whole period.  Detector conditions were
sufficiently stable throughout the livetime sample that we may assume
them constant for this analysis.  Also, the exposure was sufficiently
constant ($\chi^2$ of 75.3/95) in sidereal time to be taken as uniform
for this period, as shown in Fig.~\ref{fig:sidereal}.

The final SK-I UGM data set contains 1892 through-going muons and 467
stopping muons.

\subsection{Event Background}
\label{sec:background}
In the energy range considered, the non-astrophysical background to our
studies is dominated by atmospheric neutrinos.  To understand if the \sk
data also contain some small contribution from astrophysical sources,
source-free reference data are generated, used to test search
methods, and compared with the real data set. Two methods were
used: Monte Carlo simulations of atmospheric neutrino events, including
detailed simulation of detector effects; and the process of
``bootstrapping'' the real event data~\citep{Simpson86}.

In the bootstrap method, simulated event directions are generated by
randomly selecting uncorrelated event times and directions from the real
event database, and combining these time/direction pairs into a new
simulated ``event'' which will project back to a different point on the
sky than either of the real events from with the time and direction were
drawn.  This procedure is repeated until a statistically adequate sample
is obtained. One thousand ``fake'' sky maps with the same total number
of UGMs as the database were generated to obtain the background
estimates. Because real event times and directions are used, any
systematics due to UGM angular distribution, detector angular
acceptance, neutrino oscillation effects, live time variations between
sky locations, etc., will be automatically included in the background
estimate, while any correlations present in the data will be destroyed.
However, given the limited statistics of the real data in any given
declination band (shuffling in time only changes the event's right
ascension), these backgrounds can be too granular, introducing
non-Poissonian effects.

Analyses that did not employ the bootstrap method instead used Monte
Carlo data to simulate the background.  In order to compare expected
background distributions with the 2359~UGM events observed, 40 years of
simulated atmospheric neutrino data (14610.0 equivalent live days) were
generated using Monte Carlo techniques~\citep{Desai04}.  Because the
Monte Carlo events are generated without associated times, random times 
within the same livetime range as the dataset were assigned to Monte 
Carlo events to generate fake data samples.


\section{Time-Integrated Searches}
\label{sec:DCsearches}

This section describes searches for possible astrophysical point sources
of neutrino-induced UGMs looking just at the integrated
data set, to see if there is a steady-state excess of events from some
point in the sky.  Previously the data were examined for evidence of such
an excess from the centers of the Earth, Sun, and Galaxy to set limits
on possible WIMP annihilation~\citep{Desai04}.  Time-dependent searches
are described in \S~\ref{sec:timedependent}.

\subsection{Poisson Probability Sky Maps}
\label{sec:skymaps}

Several independent studies were performed on the SK-I dataset.  The
intuitive starting place for a point source search is a simple binned
sky map.  To generate such a sky map, the portion of the sky visible at
\sk for UGMs ({\it i.e.,} below about $54^{\circ}$ declination) is
divided into bins, the number of events in each is then counted and a
probability for finding the number of events due to background is
generated for each bin.  One study used $4^{\circ} \times 4^{\circ}$
square bins which evenly tiled the visible sky in an igloo
pixelisation~\citep{1998astro.ph..6374C}.  This size is somewhat smaller
than the more optimal $4^{\circ}$ half-angle cone discussed in
\S~\ref{sec:fluxlimits}, but is the closest size igloo structure able to
evenly map onto the sky.  This square-bin study was performed four
times, with bin center shifts of $2^{\circ}$ in right ascension and
declination applied to account for potential sources near boundaries.

Another study employed overlapping circular skymap bins, improving the
sensitivity to potential sources near bin boundaries and better matching
the radial rather than rectangular expected shape of a source
distribution.  
Though better able to avoid the
potential effects of bin edges masking a real source, this oversampling
by overlapping bins introduces non-Poissonian statistical effects which
complicate error estimation.

\begin{figure}[htbp]
  \begin{center}
\ifthenelse{\boolean{ApJnames}}
     {\plotone{f2-color.eps}}  
     {\plotone{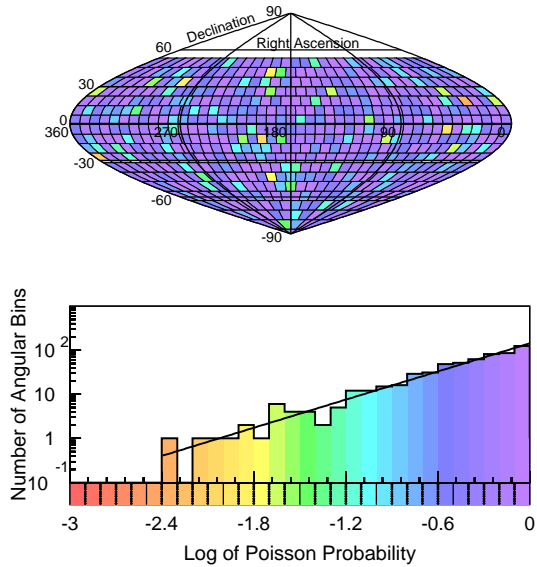}}
  \end{center}
  \caption{Probability of the observed number of neutrinos in a skymap bin
    being a statistical fluctuation, for tiled $4^\circ\times4^\circ$
    bins, shown on the sky (top) and as a histogram (bottom).  
The distribution of probabilities
    (straight-line fit in bottom figure, slope $=1.06\pm0.04$ and
    reduced $\chi^2=$ 0.83) is consistent with expectation
    for a Poisson distribution.}
  \label{poisson_map_tiledbins}
\end{figure}

Both the square and circular bin studies used bootstrap estimates of the
background noise in each bin.  They both produced cumulative Poisson
probability sky maps, an example of which is shown in
Fig.~\ref{poisson_map_tiledbins} for one of the square bin searches.  A
point source would appear here as one or more bins with an anomalously
high count relative to Poisson probability. However, in both searches the
data were consistent with the expected background, as shown in the
histogram of observed Poissonian probabilities in the lower half of the
figure.  A sample drawn from a Poisson distribution should fit a power law
of unit slope. The fit to the observed probability distribution has a
slope of $1.06\pm0.04$, with a reduced $\chi^2 =$~0.83.  No
significantly improbable bins appeared in any of the sky maps described
above.

\subsection{Clustering Analysis}
\label{sec:clustering}

Binned searches, such as the sky maps described above, treat the center
of each bin as a test direction in which to look for a source,
decreasing the sensitivity if a source really lies near a bin
boundary.  Another approach is to look for statistically significant
clustering of events on the sky map in a binning-free way.  One such
method~\citep{Oyama89,Ambro01} involves using each event as
a test direction. By examining the number of events in a cone around the
test direction and comparing to the distribution produced by the
background, one can look for regions of excess without boundary effects.

\begin{figure}[htbp] \begin{center}
\ifthenelse{\boolean{ApJnames}}
 {\plotone{f3.eps}}
 {\plotone{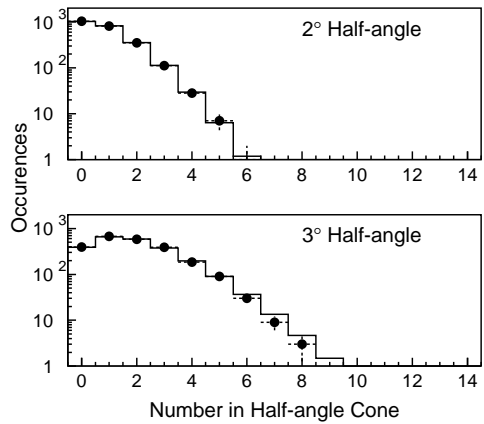}}
 \caption{Search for clustering around event directions.  Plots show the
number of occurrences of UGM clusters with multiplicity given on the
horizontal axis.  Upper plot is for $2^\circ$ half-angle search cones, and
lower plot is for $3^\circ$ half-angle cones. Data (points with error
bars) are consistent with Monte Carlo expectation for background
(histograms).}

\label{clustering}
\end{center}
\end{figure}

This cluster-search algorithm was used on the \sk UGM data.  The
expected background was calculated by dividing a Monte Carlo data sample
equivalent to 40 years of \sk operation into eight sets of 2359
simulated events to match the size of the actual data set.  Using each
event as a test direction, the number of events in a cone of half-angle
$2^{\circ}$ around each test direction were counted. The study was also
done with an expanded cone of $3^{\circ}$.  The distributions produced
by the data show no significant deviation from those produced by the the
average of the eight Monte Carlo-generated background estimates,
Fig.~\ref{clustering}.  The $2^{\circ}$ search has a comparable
solid-angle to the binned all-sky surveys in \S~\ref{sec:skymaps}, and
the $3^{\circ}$ matches the cones used for flux limits in
\S~\ref{sec:fluxlimits}.

\subsection{Specific Flux Limits}
\label{sec:fluxlimits}

None of the all-sky search methods described previously found a
statistically significant excess of neutrino-induced muons over
expectation from the atmospheric neutrino background. We thus conclude
that \sk has not detected any sources of astrophysical neutrinos.

Upper limits on the possible astrophysical neutrino-induced muon flux can
be then be set for locations on the celestial sphere with
$\delta<54^\circ$.  \sk's northern location limits the exposure of the
experiment to mainly sources in the Southern hemisphere.  

Flux limits have been calculated for a list of high-energy astrophysical
objects which have been described in the literature as potential
neutrino sources. For convenience of comparison, we started with the
same list of sources analyzed by MACRO~\citep{Ambro01}. To these we
added the catalog of SGR's (discussed in \S~\ref{sec:sgrs}) and other
potential steady-state sources of TeV neutrinos such as LS5039~\citep{Ahar05},
WR20a~\citep{Bednarek05}, 1ES~1959+650~\citep{Reimer05}, B1509-58,
B1706-44, and B1823-13~\citep{Link05}. For the flux estimates, UGMs
within $3^\circ$ of the source were selected as the signal. An angular window of  $3^{\circ}$  contains about
52\%  of the signal for an  atmospheric neutrino spectrum and 92\% of the signal for
a $\frac{1}{E^2}$ neutrino spectrum. 
 To evaluate
the expected background, we used the atmospheric neutrino Monte-Carlo 
events.
Each Monte Carlo event was assigned a time from that of the observed UGMs, in order to match the 
livetime distribution of the observed events. Then using this time and observed zenith and azimuthal 
angle,  we can obtain 
the right ascension and declination for every Monte Carlo event,  from which the angular separation
between this event and any celestial object can be estimated.  

The numbers of
signal and background were used to compute a 90\%~C.L.  number of UGMs
attributable to a background fluctuation using the procedure described in \citet{Caso98}, and that 
number was used to
compute a flux in $\rm{cm}^{-2}\rm{s}^{-1}$ as discussed
in~\citet{Fukuda99}.  The results are summarized in
Table~\ref{tab:pointsourcesummary}.

To minimize {\it a priori} issues, the approach of this paper has been
to first do the all-sky search to see if anything anywhere is
statistically significant (\S~\ref{sec:skymaps}).  Since there was
not, UGM flux limits can be placed at specific locations of interest
(Table~\ref{tab:pointsourcesummary}).

\subsection{Diffuse Flux from Galactic Plane}
\label{sec:diffuse}

Galactic and extra-galactic cosmic rays can interact in 
with the interstellar medium (ISM) of our galaxy, producing a
cascade of high energy particles, including neutrinos. There have been
various estimates of this flux in the literature~\citep{Stecker79,Berezinsky93,Ingelman96,Candia05}. 

\begin{figure}[htbp]
  \begin{center}
\ifthenelse{\boolean{ApJnames}}
    {\plotone{f4.eps}}
    {\plotone{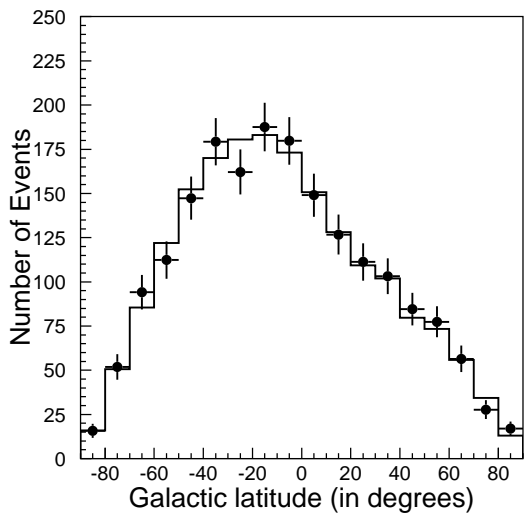}}
  \end{center}
  \caption{ \label{ismflux} Galactic latitude distribution of upward
    through-going muon dataset compared with Monte Carlo expectation from
    atmospheric neutrinos.  There is no significant excess near galactic
    latitude of $0^{\circ}$, which would be a signal of neutrinos being produced by cosmic ray
    interactions with the ISM.}
\end{figure}

The density of hydrogen in our galaxy scales with height approximately as
$\rho\propto e^{-h/h_0}$ with $h_0=0.26~\rm{kpc}$. Thus, most of the signal would
come from the galactic plane. To search for UGMs from cosmic rays
interacting with the galactic ISM, we first looked for an excess within 
$\pm 10^{\circ}$ around the Galactic plane.
Since the flux of ISM-induced neutrinos is expected to be comparable to
that of atmospheric neutrinos only at energies above
100~GeV~\citep{Ingelman96}, we performed this search with only upward through-going muons.  
The galactic latitude distribution of upward through-going muons is 
shown in
Fig.~\ref{ismflux}.  Compared 
with the same
distribution for the Monte Carlo data, there was no statistically
significant excess within $\pm 10^{\circ}$ of galactic latitude of  zero degrees. Thus at the 
90\% CL, after subtraction of expected atmospheric neutrino background, 
less than 21  upward through-going 
muons  observed at \sk are from 
cosmic rays interacting with the ISM. This flux limit is consistent with the predictions from the model 
by~\citet{Berezinsky93}, which when extrapolated to the size of \sk predicts about  2.3 events. 
We also note that recently a similar search for high energy neutrinos from the galactic plane has been 
performed with the AMANDA-II 
detector~\citep{Kelley05}.

\section{Time-Dependent Searches}
\label{sec:timedependent}

In addition to direction, we also have timing information to aid in the
search for point sources.  This section describes searches for possible
astrophysical point sources of neutrino-induced upward going muons which
also consider the arrival times of the events in question, to be more
sensitive to potentially transient sources than the ``DC''-style
searches described in \S~\ref{sec:DCsearches}.  For these searches the
timing cut used helps reduce the atmospheric neutrino background,
allowing the use of a larger angular cone size ($5^\circ$ half-angle)
as compared to time-integrated searches in order to maximize any possible
signal.

\subsection{Triggered space-time correlation searches}
\label{sec:triggered}

If a potential source is known to be transient or come in bursts from
electromagnetic observations, the neutrino data can be examined in
coincidence with these external ``trigger'' observations.  The SK-I data
set has already been searched for time coincidences with BATSE GRB
triggers.  ~\citep{Fukuda02}. 

\subsubsection{Markarian 501 Space-Time Correlation Study}
\label{sec:mkr501}

The flaring episode of Markarian 501~\citep{Weekes00} between February
and October of 1997 was used to search for possible time coincidences
with the SK-I UGM data.  MRK~501 is a BL~Lac object and a known TeV
$\gamma$-ray source.

A $5^\circ$ half-angle cone centered on MRK~501 was used to count events
recorded during periods of heightened activity, which accounted for
13.5\% of SK-I live time. Another 67.5\% was
logged definitely before or after the flaring event. The remainder of 
the livetime corresponded to periods when MRK~501 was not observable by 
air Cherenkov telescopes, so its level of activity could not be 
determined and was not used in this analysis.

One UGM was observed within this half-angle cone and during the 
flaring period. The estimated background which was estimated using the
Monte-carlo data was 1.6. Given this background, at least seven events would be required 
in order to claim a  $3\sigma$ signal detection. Therefore, the SK-I dataset provides no evidence for
MRK~501 acting as a high-energy neutrino point source during its 1997 gamma ray outbursts.


\subsubsection{SGRs Space-Time Correlation Study}
\label{sec:sgrs}

Soft Gamma Repeaters (SGRs) are a small class (four confirmed and three
tentative candidates) of high energy transient sources which emit bursts of X-rays
and gamma-rays with durations lasting from 0.1 to 500 sec, and are located
in the galactic plane~\citep{Hurley99,Woods04}. Originally confused with GRBs,
their burst properties differ from those of GRBs in two respects:  SGR
bursts are repetitive, and have a softer spectral energy distribution. On
rare occasions, SGRs have also emitted giant flares with hard spectra up
to MeV energies. The total photon fluence received at earth varies from
$10^{-9} \rm{erg/cm}^2$ to almost $10^{-3} \rm{erg/cm}^2$. The number of
bursts as a function of energy is given by $\frac{dN}{dE} \approx
E^{-1.7}$. The currently favored model for burst emission from SGRs~\citep{thomp95} is 
magnetic field decay during crustquakes in neutron stars with 
$B~>~10^{14}$~G (which are called ``magnetar''). The flux of high-energy 
neutrinos during
a giant flare has already been estimated for the SGR 1806-20 flare on December 27, 2004,
which was during SK-II~\citep{Ioka05,Halzen05}. There are also models for 
steady-state emission
of high energy neutrinos from selected SGRs~\citep{Zhang03}.

\begin{figure}[htbp]
  \begin{center}
\ifthenelse{\boolean{ApJnames}}
 {\plotone{f5.eps}}
 {\plotone{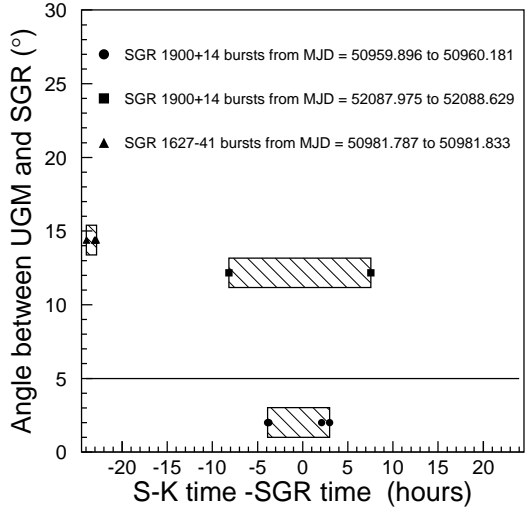}}
 \caption{\label{sgrupmutimecorrelation} Coincidences between SGRs and
   \sk upward muons in a $\Delta T = \pm 1 $~day and 
$\Delta \theta = 30^{\circ}$ window, where $\Delta T$ and $\Delta \theta$
indicate the arrival time difference and angle respectively between the
Super-K UGM and SGR.  Each hatched box represents a correlation between
 an UGM  and a group of SGR bursts which occur within a short time interval.
 The vertical mid-point of each box gives the direction between the
 observed UGM and the SGR, and the vertical size of the box is
 $2^{\circ}$, approximately the angular resolution for UGMs.  Each symbol
 within a box gives the time of an SGR outburst relative to the UGM.
 The width of each box is the time interval between the first and last
 correlated SGR bursts.  Within an angular search window of $\Delta
 \theta = 5^{\circ}$ we found one UGM within hours of four different
 bursts from SGR~1900+14 (the bottom box) between 1998 May 26,
 21:30:29~UT and 1998 May 27, 04:22:01~UT.}
  \end{center}
\end{figure}

We searched the SK-I data for space-time correlations between UGMs and
SGR bursts in an analysis similar to that used previously for
GRBs~\citep{Fukuda02}. SGR burst times were obtained from the
Interplanetary Network (IPN) catalog (K.~Hurley, private communication).
During SK-I only four SGRs were observed to emit bursts. These are
SGR~1806-20 (33 bursts), SGR~1900+14 (67 bursts), SGR~1627-41 (51
bursts), and SGR~1801-23 (2 bursts). We searched for coincidences within
a window of $\Delta T = \pm 1$~day and $\Delta \theta = 5^{\circ}$. For
this time window size, the total number of bursts from SGRs occurring in
separate time bins was 74. The space-time coincidences found between
SGRs and UGMs are shown in Fig.~\ref{sgrupmutimecorrelation}.

No coincidences were found from SGR~1806-20, SGR~1627-41 and SGR~1801-23
and SGR~1627-41. One UGM was found within $2^{\circ}$ of SGR~1900+14 and
within several hours of four bursts starting 1998 May 26, 21:30:29~UT
and ending 1998 May 27, 04:22:01~UT. The difference in times between
the bursts and the UGM were -3.88, -3.75, 2.08, and 2.97 hours.

The expected background in a $\Delta T = \pm 1$~day window calculated
using the Monte Carlo data is 0.013 events. Given that 74
different SGR bursts were examined, the total background becomes 0.96.  
Therefore the UGM found coincident with SGR~1900+14 is consistent with
the background and not statistically significant.

\subsection{Untriggered Searches for Burst Sources}
\label{sec:untriggered}

In the space-time correlation studies for UGMs described above, the
reported times and sky coordinates of astrophysical photon signals from
other detectors were used as triggers to look for correlations in the SK-I
neutrino data. We also looked for bursts of UGMs from astrophysical
sources independent of photon observations, in order to
detect transient sources of high energy neutrinos for which there may be
no electromagnetic counterpart. The method used 
was to regard each
observed UGM as a trigger itself, and look for other upward muon
events arriving within a time window of one hour and an angular window of
$5^{\circ}$. 

\begin{figure}[htbp]
  \begin{center}
\ifthenelse{\boolean{ApJnames}}
 {\plotone{f6.eps}}
 {\plotone{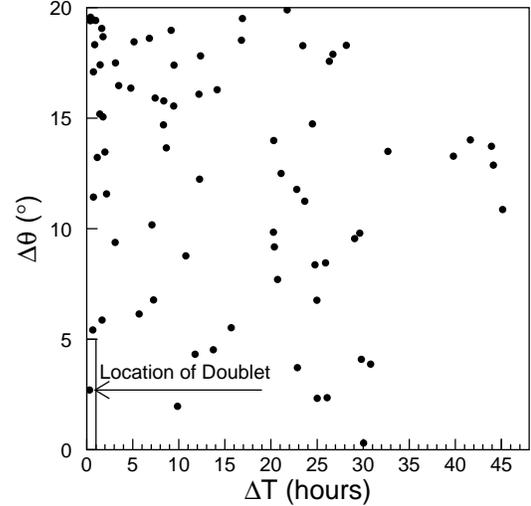}}
 \caption{\label{untriggered} A scatter plot of the relative angle ($\Delta \theta$) 
 vs  time-difference ($\Delta$ T) between consecutive UGMs (when arranged in chronological
order) in SK-I, within a window of $\Delta$ T =  2 days and $\Delta \theta = 20^{\circ}$.  One 
UGM doublet was found in the signal window of $\Delta$ T = 1 hour and $\Delta \theta = 5^{\circ}$. 
}
  \end{center}
\end{figure}

As we see from Fig.~\ref{untriggered}, using the entire SK-I dataset, one doublet was found to meet 
this 
criteria.  Two UGMs separated by $2.7^{\circ}$ arrived within 19
minutes.  The projected right ascension and declination for these two 
events were 
$7^{\rm h} 24^{\rm m} 0^{\rm s}$, $16^{\circ} 18$' and $7^{\rm h} 32^{\rm
  m} 24^{\rm s}$, $14^{\circ} 30$' respectively.

The arrival times of the events were Modified Julian Date = 50927.64412 and 50927.65757, or
1998 April 24, 15:27:32 and 15:46:54~UT.

 The expected rate of such coincidences is $2 \times 10^{-5}$ per UGM.
When multiplied by the number of trial factors
(number of UGMs minus one), the probability that this doublet comes from a
background fluctuation is about 5\%, which is not statistically
significant.  Thus we conclude that we do not see any evidence for bursts
of UGMs from an astrophysical source in the SK-I dataset.


\section{Summary}
\label{sec:summary}

Several independent methods for finding astrophysical sources of high
energy neutrinos have been applied to the SK-I (1996 to 2001)
upward-going muon data using both directional and temporal information.
None have uncovered a statistically significant excess above the
atmospheric neutrino background as estimated from bootstrapped data and
Monte Carlo simulations.  Thus, upper limits were set on the
upward-going muon flux.  The 90\% limits from astronomical sources which
are located in southern hemisphere and always under the horizon for \sk
are $(1 \sim 4)\times 10^{-15}\rm{cm}^{-2}\rm{s}^{-1}$.

\acknowledgements 
We gratefully acknowledge the cooperation of the Kamioka Mining and
Smelting Company.  The Super-Kamiokande experiment has been built and
operated from funding by the Japanese Ministry of Education, Culture,
Sports, Science and Technology, the United States Department of Energy,
and the U.S. National Science Foundation, with support for individual
researchers from Research Corporation's Cottrell College Science Award.
We would like to thank Kevin Hurley for providing us the IPN data on SGRs.  






\begin{deluxetable}{lccccc}
\tablecaption{\label{tab:pointsourcesummary} The allsky surveys showed no
significant point source, so these data are used to compute 90\% flux
upper limits within a $3^\circ$ half-angle cone of the given equatorial
coordinates of interest to astrophysical modeling. }  
\tablewidth{0pt}
\tablehead{
\textbf{Source} & 
\textbf{RA} & 
\textbf{Dec} & 
\textbf{\# events} & 
\textbf{Bkgnd} & 
\textbf{$90\% ~c.l.~\mu$ flux limits} \\
\textbf{} & \textbf{($^{\circ}$)} & \textbf{($^{\circ}$)} & & &  
\textbf{($10^{-15} cm^{-2} s^{-1}$)}}
\startdata
SMC X-1 \tablenotemark{1}  & 19.3  &-73.4 & 0 &  3.0 &  1.25 \\
LMC X-2 \tablenotemark{1} & 80.1  & -72.0 & 5 & 3.1 & 3.49 \\
SN 1987A\tablenotemark{2}  & 83.9  &-69.3 &  2 & 3.2 & 1.9 \\
LMC X-4\tablenotemark{1}  & 83.2 & -66.4 & 3 &3.3 & 2.3 \\
GX301.2 \tablenotemark{1} & 186.7  & -62.8 & 5 &  3.5 & 3.35 \\
Cen X-5 \tablenotemark{1} &  177.0  & -62.2 & 4 &  3.65 & 2.73\\
GX 304-1 \tablenotemark{1} & 195.3  & -61.6 & 4 & 2.1 & 3.33 \\
Cen X-3 \tablenotemark{1} & 170.3  & -60.5&  0 &  3.2 & 1.27 \\
Cir X-1 \tablenotemark{1}&  230.2  &-57.2& 5 & 3.3 & 3.46 \\
2U 1637-53 \tablenotemark{1} &  250.2  &-53.7 & 3  &  2.9  & 1.99 \\
4U 1608-522 \tablenotemark{1} &  243.2  &-52.4 & 5 &  3.6  & 3.10 \\
GX 339.4 \tablenotemark{1}  & 255.7  & -48.8&    1  &  2.4  & 2.02 \\
Vela \tablenotemark{3} & 128.8  & -45.2 &        1  &  2.7  & 1.77 \\
GX 346-7 \tablenotemark{1} & 264.7  & -44.4& 4  &  2.4  &  4.29 \\
AR X-1 0 \tablenotemark{1} &256.6 & -43.0 &  2  &   2.9  & 2.66 \\
SN 1006 \tablenotemark{2} & 225.7  & -41.9&  3  &  2.4  &  3.65 \\
Vela X-1 \tablenotemark{1} &  135.5  & -40.6&   2  &  1.7  & 3.23 \\
2U 1700-37 \tablenotemark{4} & 226.0  & -37.8& 0  & 2.6&1.90 \\
SGR X-4 \tablenotemark{1} &  275.9  & -30.4 & 4 &  2.0  & 4.44  \\
L10 \tablenotemark{5}  & 269.4  & -30.1&     3  & 2.0  & 4.41 \\
GX 1+4 \tablenotemark{1} & 263.0   & -24.7 & 2 &  1.5  & 2.89 \\
SN 1604 \tablenotemark{2}  & 262.7  & -21.4 & 3 & 1.8  & 3.79  \\
GX 9.9 \tablenotemark{1}  & 262.9  & -17.0 & 2  &  1.6 & 4.00 \\
Sco X-1 \tablenotemark{1} & 245.0  &  -15.6 &3 &1.25 & 5.43 \\
Aqr X-1 \tablenotemark{6}  & 310.0  & -0.9 &0 & 1.7 & 2.5 \\
4U 336+01 \tablenotemark{1} & 54.2   & 0.6 &2  & 1.5 & 4.54 \\
Aql  X-1  \tablenotemark{1}  & 287.8  &  0.6 & 0  & 1.7 & 2.53 \\
2U 1907+02  \tablenotemark{7} & 286.8  &  2.3 & 1  & 1.9 & 3.35 \\
Ser X-1  \tablenotemark{1} &  280.0  &  5.0 & 3   & 1.6 & 4.61 \\
SS433  \tablenotemark{4} & 287.9  & 5.0 &   2   & 1.5 &  3.52 \\
2U 0613+09 \tablenotemark{1} & 93.4  & 9.1 &        0 & 1.3  & 2.7 \\
Geminga  \tablenotemark{3}  & 98.4  & 17.8&         0 & 1.3  & 2.94 \\
Crab   \tablenotemark{3}  & 83.6  & 22.0 &   0  & 2.0  & 3.09 \\
2U 035+30 \tablenotemark{1} & 58.8  & 31.0 & 3  & 1.1  & 6.78 \\
Cyg X-1   \tablenotemark{1}  & 299.5  & 35.2 & 3  & 1.6  & 6.84 \\
Her X-1 \tablenotemark{1}  &254.4  &35.3  &  2  & 1.5  & 6.94 \\
Mrk 421 \tablenotemark{8} &  166.11  &38.2 & 2  & 1.0  & 5.98 \\
Cyg X-2 \tablenotemark{1}  & 326.2  & 38.3 & 2   &  0.8  & 6.07 \\
Mrk 501 \tablenotemark{8} &  253.5  & 39.8 & 0  & 1.0  & 4.39 \\
Cyg X-3 \tablenotemark{9} & 308.1  & 40.9 &  1  & 0.8  & 6.66 \\
Per X-1 \tablenotemark{10} & 49.6    & 41.5 & 0  & 1.2  & 4.66 \\ 
SGR 1806 \tablenotemark{11} & 271.5 & -20.41 & 1 & 2.3 & 2.82 \\ 
SGR 1900 \tablenotemark{11} & 286.8 & 9.3 & 1 & 1.6 & 3.62\\ 
SGR 1627 \tablenotemark{11} & 246.8 & -41.0 & 2 & 1.9 & 3.11 \\ 
SGR 1801 \tablenotemark{11} & 270.3 & -23.0 & 1 & 2.2 & 2.78 \\ 
SGR 0525 \tablenotemark{11} & 81.2 & -66.0 & 5 & 2.75 & 3.64 \\ 
LS 5039 \tablenotemark{1}  & 276.5 & -14.8 & 1 & 1.2 & 3.18 \\
WR 20a \tablenotemark{9} & 156 & -57.8 & 3 & 2.7 & 2.51 \\
1ES 1959+650 \tablenotemark{11} & 300 & 6.8 & 3 & 1.3 & 6.31 \\
B1509-58 \tablenotemark{3} & 227.2 & -58 & 6 & 2.8 & 3.67 \\
B1706-44 \tablenotemark{3} & 257.2 & -44.5 & 1 & 2.9 & 2.14 \\
B1823-13 \tablenotemark{3} & 275.6 & -13.2 & 3 & 1.35 & 5.43 \\
\enddata 

\tablenotetext{1}{X-Ray Binary}
\tablenotetext{2}{SN Remnant}
\tablenotetext{3}{Pulsar}
\tablenotetext{4}{Emission Line Star}
\tablenotetext{5}{Dark Nebula}
\tablenotetext{6}{Cataclysmic binary}
\tablenotetext{7}{Symbiotic Star}
\tablenotetext{8}{Blazar}
\tablenotetext{9}{Wolf-Rayet Star}
\tablenotetext{10}{Cluster}
\tablenotetext{11}{SGR}
\end{deluxetable}

\end{document}